# Personalized brain network models for assessing structure-function relationships

Kanika Bansal[1,2,3], Johan Nakuci[4], and Sarah Feldt Muldoon[1,4,5,*]

[1] Mathematics Department, University at Buffalo – SUNY, Buffalo, NY 14260
[2] Human Sciences, US Army Research Laboratory, Aberdeen Proving Grounds, MD 21005
[3] Department of Biomedical Engineering, Columbia University, New York, NY 10027
[4] Neuroscience Program, University at Buffalo – SUNY, Buffalo, NY 14260
[5] CDSE Program, University at Buffalo – SUNY, Buffalo, NY 1260

* Corresponding author e-mail: smuldoon@buffalo.edu

## Abstract

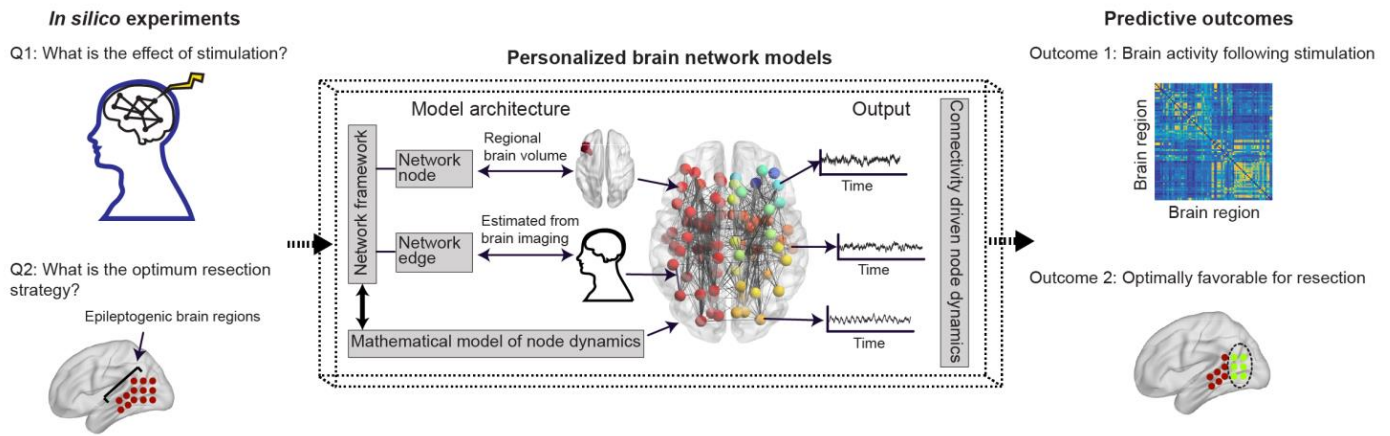

Many recent efforts in computational modeling of macro-scale brain dynamics have begun to take a data-driven approach by incorporating structural and/or functional information derived from subject data. Here, we discuss recent work using personalized brain network models to study structure-function relationships in human brains. We describe the steps necessary to build such models and show how this computational approach can provide previously unobtainable information through the ability to perform virtual experiments. Finally, we present examples of how personalized brain network models can be used to gain insight into the effects of local stimulation and improve surgical outcomes in epilepsy.

# Introduction

The brain is an inherently dynamical system, driven by an underlying complex network of connections, and much work has focused on the ability to relate brain activity and function to the underlying structure [1]. Understanding this important link has been a key factor in the development of Network Neuroscience – a rapidly developing field that relies on complex network theory to model and study the brain across multiple scales and modalities of interactions [2]. In this framework, network nodes are chosen depending upon the scale of interest and scientific question, and could range from neurons to brain regions. Network edges can represent structural connections (anatomical links; structural connectivity) or functional relationships (statistical relationships; functional connectivity) [3,4]. When using network theory to model the brain, many important questions can be asked. What is the relationship between structural and functional connectivity? Do structure-function relationships change over task, time, or disease state? How sensitive are the observed patterns of brain activity to small differences in the underlying structural connectivity?

Studies have shown that while certain features of brain network structure are conserved across individuals, differences in network structure can be observed across people [5-7]. Individual differences in human task performance [8,9] and differences between healthy and diseased individuals [10,11] have also been linked to differences in the underlying structural connectivity of the brain. These findings have motivated the formulation of data-driven computational models of brain activity (see Box 1). These personalized brain network models (BNM) combine an individual's structural connectivity with mathematical equations of neuronal activity in order to produce a subject-specific simulation of spatiotemporal brain activity. Due to recent advances in non-invasive imaging techniques to measure macro-scale structural connectivity of human brains [12,13], such models have gained popularity to study large-scale brain dynamics. Mathematical equations are used to simulate the dynamics of each node (brain region) and are coupled through the subject-specific structural connectivity.

These computational models are sensitive to the underlying network structure [14], and offer many advantages when investigating structure-function relationships. For example, one can perform *in silico* experiments that perturb the underlying brain structure such as lesioning (removing edges [15,16]) or resection (removing nodes [17,18]) and investigate the effects of such perturbations on simulated brain activity. Alternatively, one can impact local brain dynamics through modifications to the mathematical equations such as applying stimulation or modifying brain excitability and study the effects of these local perturbations on global brain function [14,19]. Importantly, due to the specificity of the model to a given individual, one can study the differential impact of similar perturbations across a cohort of individuals. Thus, this approach has the potential to lead to the development of personalized treatment strategies to combat disease or enhance human performance [20].

In this review, we summarize the basic steps involved in creating personalized BNM and provide examples from recent studies within the last 2-3 years that have used this methodology to gain insight in to brain structure-function relationships. We particularly highlight applications of this approach that study effects of regional brain stimulation on global brain dynamics or use computational brain models to predict surgical outcomes in epilepsy.

---

**Box 1: Personalized brain network models (BNM)**

"Everything should be made as simple as possible, but not simpler" - Albert Einstein.

Network neuroscience seeks to understand the organization of the brain using tools from complex network theory, applied across multiple scales and modalities. Given the ongoing experimental advances in non-invasive recording techniques, it is now possible to combine high quality structural brain data with neurophysiological information to create data-driven computational models of brain activity. These personalized BNM simulate brain dynamics using biologically inspired mathematical equations that model regional activity and are coupled through the observed brain structure. Incorporating personal data into the structure and dynamics of the model involves making multiple assumptions and choices that are driven by the question at hand. The flexibility associated with the model design makes it useful for performing *in silico* experiments across a diverse range of applications, but also implies that one must be cautious when interpreting model predictions and/or making generalizations.

**Applications:**
Personalized BNM can
- be tuned to produce dynamics that mimic the resting state activity patterns.
- predict the effect of targeted stimulation.
- be perturbed to study the impact of brain lesions.
- provide seizure onset probabilities and inform surgical outcomes.

**Limitations:**
When using macro-scale computational models, one must also keep in mind the underlying assumptions and limitations. These models
- are often optimized based on the scientific question at hand and are not always generalizable.
- do not necessarily produce waveforms that depict realistic brain activity.
- can provide predictive outcomes but might lack many neurophysiological details and/or mechanistic explanations.

# Building data-driven brain network models

Building personalized BNM involves making decisions regarding the scale of the network, the type of underlying connectivity, and the level of neurophysiological complexity of the mathematical equations that constitute the model (see Fig. 1). Generally, human brain modeling involves working at the macroscopic level where network nodes are either sensors (e.g., EEG data) or brain regions (e.g., imaging data) [21]. Brain regions are defined using a parcellation scheme that is based on one of many different available atlases [22]. Each atlas divides the brain into multiple spatial regions, but the location and total number of regions varies widely between atlases. Due to this variability, some work has investigated the impact of the choice or scale of atlas used for the parcellation. While the proper choice of scale depends on many factors, under certain sets of assumptions, it has been shown that an atlas with approximately 140 brain regions produces good agreement with experimental data [23].

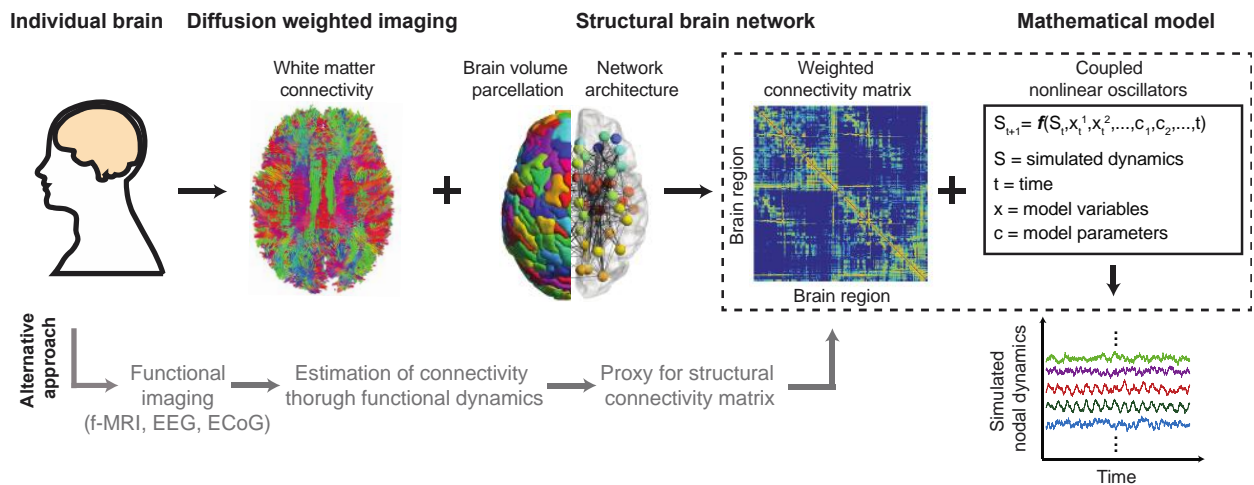

**Figure 1. Building personalized brain network models.** To construct data-driven BNM, individual brain connectivity is combined with mathematical equations within a complex networks framework. Typically, the brain is parcellated into different regions which constitute network nodes, and network connectivity is derived from diffusion weighted imaging that estimates the density of white matter tracts between brain regions. This results in a weighted connectivity matrix whose entries represent the connection strength between brain regions and is specific to a given individual. The dynamics of each brain region are simulated using biologically inspired mathematical equations that are coupled though the connectivity matrix and can also be further tuned to match a specific individual's activity patterns. Occasionally, functional connectivity is used as a proxy for structural connectivity data.

Next, one needs to determine how network edges are defined. The modeling framework is based on the assumption that subject-specific structural connectivity data is available. In such cases, the connectivity is commonly derived from estimates of white matter tracts between brain regions. The first modeling studies used a connectome derived from tract tracing studies in primates [24], but in order to model human brain dynamics, estimates of white matter tracts obtained from diffusion weighted imaging data are instead used [25]. While some studies have used structural networks that represent data averaged across a cohort of individuals [26,27], using subject-specific connectomes increases the specificity of the model [19,25,28] and is preferable if available.

In certain cases where subject-specific structural network data is not available, researchers have instead substituted functional for structural connectivity. However, it is important to remember that the mathematical assumption of the model is that the connections represent structural (not functional) coupling. While it has been shown that structural and functional networks are correlated [25,29], these two types of connectivity remain fundamentally different [30]. Nevertheless, assessing brain network model connectivity from functional data can still be shown to produce predictive results [31] and can therefore be a useful tool, but one must be careful in the interpretation of the findings.

Finally, one must choose the set of mathematical equations that represent regional brain activity. In the simplest case, Kuramoto phase oscillators (simplistic oscillators commonly used in dynamical systems theory [32]) have been used to model neural activity. However, more sophisticated approaches instead choose some biologically informed neural mass model. See the excellent review by Breakspear [33] for a detailed discussion on the choices of dynamical equations. If one's goal is to be able to accentuate the effect of the underlying structural connectivity, each brain region is generally governed by the same set of equations and same parameters. However, to more accurately model brain activity, one can modify parameters of the equations governing regional brain dynamics. This approach is particularly to model changes related to brain state such as sleep vs. wake [26,27], or disease states such as epilepsy [19], and is becoming increasingly used to create models that capture subject specific differences in both structure and dynamics.

Thus far, there is no common rule for selecting a particular type of connectivity or mathematical equations when constructing BNM. In fact, there is a substantial element of subjectivity related to the specific problem at hand, the background of the researchers, and the types of available data. While many researchers construct BNM using in-house code, in order to accommodate the variety of possible assumptions and diverse range of neuroscientific questions, a software package called The Virtual Brain (TVB) platform was recently developed [34-36]. Within this interactive platform, users can build BNM based on multiple choices of modeling assumptions, and this platform has been used to study a wide range of applications ranging from disease states [19,28,37] to stimulation [38,39] to resting state dynamics [40]. Recently, TVB has also been extended to the mouse brain which will allow researchers to compare model predictions with a much wider range of experimental outcomes [41].

## The incorporation of personal data

Constructing *personalized* BNM requires that the model is sufficiently sensitive to individual variability in the data. It has recently been shown that when modeling brain dynamics using human connectomes derived from diffusion spectrum imaging data, the inter-subject variability is greater than the variability in multiple scans of a single subject [14]. Additionally, when regional stimulation is applied to models that differ only in the structural connectivity, the resulting activity patterns differ between individuals (see Fig. 2). This suggests that despite the many known caveats of assessing structural connectivity from imaging data [12], the method is indeed sensitive to subject-specific differences in structure.

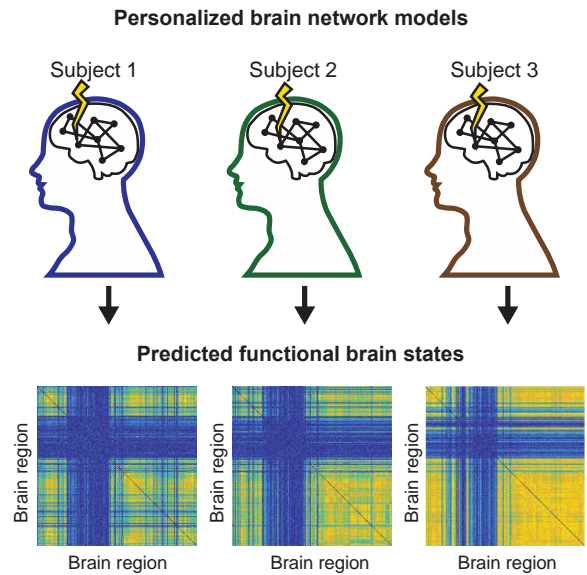

**Figure 2. Sensitivity to individual differences in underlying structure.** When performing *in silico* brain stimulation experiments, computational stimulation of the same brain region in models derived from the structural connectivity of different individuals results in visibly different functional brain states.

Further, as described above, the model incorporates both structural and dynamical information and assumptions, and each of these can be subject-specific. In general, the ability to incorporate personal data depends on the availability of such data and typically varies from study to study. For example, each of the studies presented in this review use a different set of assumptions to model both structural connections and brain dynamics. This variability in the modeling approach is natural, but also means that one must carefully examine the modeling assumptions and use of personal data when interpreting model predictions, making generalizations, or comparing studies.

## Applications: Regional brain stimulation

An advantage of building a computational model of an individual's brain is that one can then perform virtual experiments that would not be possible due to experimental/ethical constraints. For example, there is increasing interest in using non-invasive brain stimulation such as transcranial magnetic stimulation (TMS) or transcranial direct current stimulation (tDCS) as a clinical treatment option [42]. However, the mechanisms and network effects of such stimulation remain largely unknown. Personalized BNM allow one to systematically study the effects of stimulation to different brain regions in different people [14,38,39].

Muldoon et al. recently used this approach to relate patterns of activation due to regional stimulation to predictions from network control theory, showing that stimulation of brain regions with low controllability resulted in local activation patterns while stimulation of brain regions with high controllability promoted global activation [14]. Further, they found that stimulation of brain regions within the default mode network were able to produce large global effects, despite being constrained by the underlying pattern of anatomical connectivity.

In a different study, Spiegler et al. used TVB platform to perform sequential stimulation of brain regions, showing that stimulation of certain regions produced activation patterns that were consistent with known resting state networks [39]. Importantly, experiments such as these would not be possible to perform on individuals due to practical and ethical concerns of applying systematic regional stimulation to human brains. Thus, model based approaches such as those described above are essential for informing researchers about the differential effects of regional simulation.

## Applications: Informing surgical decisions in epilepsy

Personalized BNM also offer advantages when modeling the effects of diseases such as schizophrenia [43,44], Parkinson's Disease [45,46], and epilepsy [17,19,28,31,47]. For example, epilepsy is a disorder that is known to be associated with both structural and dynamical changes in the brain [48,49]. Jirsa et al. designed a framework within TVB platform to incorporate patient specific data such as the location of seizure initiation, subject-specific connectivity, and MRI lesion data into BNM [19]. They used this computational framework to successfully predict the patterns of seizure propagation in epileptic patients [28]. Further, they showed that when simulated propagation patterns identified recruited regions not monitored clinically, surgical outcomes were poorer, suggesting that the model predictions could be used to inform clinical monitoring and improve surgical outcomes.

Other work has used personalized brain models to perform virtual resection experiments in order to make predictions about which brain regions should be targeted for surgical resection [17]. Sinha et al. compared model predictions of surgical resection sites (using functional as opposed to anatomical connectivity as the structural basis of the model) with actual surgical resections and found high overlap in model predictions and surgical removal sites [31]. They also found that in cases where actual surgical outcomes were poor, the model could identify alternative potential target areas for resection.

## Conclusion

Personalized BNM are essential to probe structure-function relationships, as they allow one to perform controlled virtual experiments not otherwise possible. While these models may lack mechanistic explanations, they provide powerful predictive outcomes that have the potential to greatly influence and advance medical treatment strategies. Additionally, these models can answer fundamental questions such as how the human brain is structurally constrained to produce functional patterns of activity or what structural/dynamical features drive individual variability in performance. Future work must also focus on classifying and understanding this individual variability in brain activity and performance. As researchers are able to incorporate more personal information into model structure and dynamics, personalized BNM will increasingly serve as an important tool for assessing structure-function relationships and designing personalized medical treatment strategies.

# Acknowledgements

SFM would like to acknowledge support from the National Science Foundation (SMA-1734795). SFM and KB also acknowledge support from the Army Research Laboratory (contract numbers W911NF-10-2-0022 and W911NF-16-2-0158 respectively). The content is solely the responsibility of the authors and does not necessarily represent the official views of any of the funding agencies.

This extensive review discusses multiple different options and associated assumptions for modeling large-scale brain dynamics using oscillator models with variable levels of biophysical realism. The paper serves as a basic introduction to dynamical systems and is accessible to readers with limited prior knowledge of dynamical systems theory.